\documentclass[apj]{emulateapj}
\newcommand{\mdeg}{\ensuremath{^{\circ}}}
\usepackage{graphicx}
\usepackage{subfigure}
\usepackage{amsmath}
\usepackage{amssymb}
\usepackage{bm}
\usepackage{url}
\shorttitle{}
\shortauthors{Yogesh Maan}
\submitted{}
\journalinfo{Accepted for publication in The Astrophysical Journal}
\begin{document}
\title{Discovery of Low DM Fast Radio Transients: Geminga Pulsar Caught in the Act}
\author{Yogesh Maan\altaffilmark{1}}
\affil{National Centre for Radio Astrophysics, Pune 411007, India}
\email{ymaan@ncra.tifr.res.in}
\altaffiltext{1}{Raman Research Institute, Bangalore 560080, India}
\begin{abstract}
We report discovery of several energetic radio bursts at 34~MHz, using the
Gauribidanur radio telescope. The radio bursts exhibit two important properties
associated with the propagation of astronomical signals through the interstellar
medium: (i) frequency dependent dispersive delays across the observing
bandwidth, and (ii) Faraday rotation of the plane of linear polarization. These
bursts sample a range of dispersion measures (DM; 1.4--3.6~pc~cm$^{-3}$), and
show DM-variation at timescales of the order of a minute. Using groups of
bursts having a consistent DM, we show that the bursts have originated from
the \emph{radio-quiet} gamma-ray pulsar Geminga. Detection of these bursts
supports the existence of occasional radio emission from Geminga.
The rare occurrence of these bursts, and the short timescale variation
in their DM (if really caused by the intervening medium or
the pulsar magnetosphere),
might provide clues as to why the pulsar has not been detected in earlier
sensitive searches. We present details of the observations and search
procedure used to discover these bursts, a detailed analysis of their
properties, and evidences of these bursts being associated with
Geminga pulsar, and discuss briefly the possible emission mechanism
of these bursts.
\end{abstract}
\section{Introduction}
Detection of fast radio transients has been key to several important
discoveries.
Discovery of several pulsars, including the first ever discovered pulsar
B1919+21 \citep{Hewish68}, has been made through the detection of their
single pulses. The Crab pulsar was also discovered through its giant
pulses \citep{SR68}. Discovery of rotating radio transients
\citep[RRATs;][]{McLaughlin06} --- a category of pulsars that are
highly intermittent in their emission --- would not have been possible
without a systematic search for transients. More recently, a new class
of bright dispersed pulses, called fast radio bursts (FRBs), has been
discovered \citep{Lorimer07,Keane12,Thornton13}. Most of the FRBs have
been discovered far from the Galactic plane, and yet at anomalously high
dispersion measures (DMs; in the range 375--1630~pc~cm$^{-3}$). The
excess in DM is assumed to be contributed by the extra-galactic medium,
and hence, most of the FRBs are considered to be of extra-galactic origin.
The narrow pulse-widths (of the order of a few milliseconds) and high brightness
temperatures of the FRBs indicate towards coherent emission processes.
Although several possible progenitors have been proposed (e.g.,
merging binary white dwarf systems \citep{KIM13}, collapsing supramassive
neutron stars \citep{FR14}, hyperflares from extra-galactic magnetars
\citep{PP07}, binary neutron star mergers \citep{Totani13}, synchrotron
maser emission from relativistic, magnetized shocks \citep{Lyubarsky14},
collisions between axion stars and neutron stars \citep{Iwazaki15}, radio
waves from pulsar companions \citep{MZ14}), the exact physical nature
of the FRBs still remains unknown and awaits for further clues from future
detections.
\par
The aforementioned discoveries, especially that of the FRBs, have
enormously boosted the searches for fast transients across a large span
of radio wavelengths. However, such explorations are still very limited
in the low frequency part of the radio spectrum ($\lesssim 100$~MHz).
Some of the transient sources at these wavelengths include sporadic emission
from known and unknown pulsars and RRATs, and radio flares from active stars
and planets (like Jovian bursts).
Recent detection of transient radio signals from the gamma-ray pulsar
J1732$-$3131 \citep{MAD12} demonstrated the potential scope of low
frequency searches in detecting the low frequency counterparts of
\emph{radio-quiet} gamma-ray pulsars. The discovery of several bright
dispersed pulses presented in this article are most
likely associated with the first ever discovered {radio-quiet}
gamma-ray pulsar J0633+1746 --- also known as Geminga pulsar.
\par
The search for continuum as well as pulsed radio emission from Geminga has
a long history, with some claimed detections as well as very sensitive upper
flux density limits on pulsed emission. Although three research groups
from the Pushchino radio
astronomy observatory claimed low significance ($<10\sigma$) detections
of pulsed radio emission from this pulsar \citep[with flux densities in
the range 30--100~mJy at 102.5~MHz;][]{MM97,KL97,SP98}
at a DM of about 3~pc~cm$^{-3}$, several other sensitive searches, before
as well as after the above claimed detections, have remained unsuccessful
\citep[e.g.,][] {Seiradakis92,Ramach98,Ershov07,Coenen_thesis,MA14}.
A possible detection of the pulsar was also reported using the
Rajkot radio telescope (India) at 103~MHz \citep{Vats99}. More
recently, \citet{Malov15} have presented further, albeit still low-significance,
detections of radio pulses from this pulsar at 42, 60 and 111~MHz. Given
the low significance of all the claimed detections so far, credibility of
these detections have been questioned against the sensitive upper limits.
Detection of strong pulses presented in this paper will hopefully provide
clues why this pulsar has been detectable only occasionally.
\par
We describe our observations and search procedure
in Section 2, present the discovery and properties of several radio bursts
in Section 3, discuss possible source of the bursts and present their most
likely association with Geminga pulsar in Section 4, followed by a summary
of the findings in Section 5.
%
%
%
%
\section{Observations and Search Procedure}
The east-west (EW) arm of the Gauribidanur radio telescope \citep*{DSS89} was
used to observe the gamma-ray pulsars J0633+1746 and J0633+0632 at 34~MHz,
with a bandwidth of 1.53~MHz.
The beamwidths of the EW array are 21' and
$25\mdeg\times\sec{\rm (zenith~angle)}$ in right ascension (RA) and
declination (Dec), respectively. Sky position of the two pulsars differ by
approximately 13'' and 11.3\mdeg~in RA and Dec, respectively. Given the above
telescope beamwidth, both the pulsars could be observed simultaneously in a
single pointing\footnote{The position coordinates ([RA,Dec]) of the pulsars
J0633$+$0632 and J0633$+$1746, when precessed to the observing epoch
(2012--2013), are [06:34:26,~6.5\mdeg] and [06:34:38,~17.8\mdeg] respectively.
Both the pulsars were observed simultaneously by pointing towards the direction
[06:34:26,~10\mdeg].}.
The effective collecting area offered
by the EW arm is 12000~m$^2$ at the instrumental zenith ($+14\mdeg.1$
declination). A total of 130 observing sessions, each typically 30~minutes
long, were conducted between June~23,~2012 and April~18,~2013.
Severe radio frequency interference (RFI) made 31 of these
sessions completely unusable, and results from the remaining 99 sessions
are presented below.
First 45
observing sessions could use only $80\%$ of the total collecting area
since $10\%$ of the dipoles at each of the two far ends of the arm had
some technical problem. Results of deep searches for the periodic radio
signals from both the pulsars, using these 45 observing sessions, were
reported in \citet{MA14}. Observations towards several radio-quiet
gamma-ray pulsars were also reported by \citet{MA14}, which serve as blank
sky observations for the findings presented below. Additionally, at least one
of the two control pulsars --- B0834+06 and B1919+21 --- was observed on the
days when the target source observations were conducted.
\par
During each of the observing sessions, Nyquist sampled raw voltage sequence
is recorded with 2-bit, 4-level quantization. In the off-line processing,
the voltage time sequences are converted to filterbank format with desired
spectral and temporal resolutions.
To identify the parts of data that are contaminated by RFI,
robust mean and standard deviation are
calculated separately in the time and frequency domains. Appropriate
threshold in signal-to-noise ratio (typically around 10) is used to
identify the RFI contaminated spectral channels and time samples
separately. Identifying the bad frequency channels involves comparing
the above statistical quantities with those of a smoothed, fitted
radio spectrum (reflecting largely the instrumental response, i.e.,
the filter bandshape). The potential RFI contaminated time samples are
excluded while identifying the bad frequency channels and vice-versa.
Note that the above threshold based algorithm works well even in the
presence of pulsed radio signals from a pulsar, since (1) the intrinsic
spectra of pulsars are smooth, and
(2) the time-smearing of pulsed signals due to dispersion in the
interstellar medium is quite significant, especially at such low frequencies.
The frequency channels and time samples identified as RFI contaminated
are excluded from any further processing.
\par
The filterbank data, having typically
512 spectral channels at a resolution of $\sim 1$~ms, are searched for
fast transients in the DM range of 0--50~pc~cm$^{-3}$. Briefly, the search
for fast transients involves dedispersing the filterbank data at several
optimally spaced trial DMs, and searching the individual dedispersed
time series for candidate pulses above a specified detection threshold. For
an optimum detection, a search across the pulse-width is also conducted in
each of the dedispersed time series. For more details of the search procedure
as well as those of the acquisition setup, the Gauribidanur radio telescope,
and the pre-search data processing, we refer the reader to \citet{MAD12}
and \citet{ythesis,MA14}.
%
\section{Discovery of Radio Transients and Their Properties}
Our search for fast radio transients resulted in detection of
bright radio pulses from the observing sessions conducted on
July~13,~2012 and July~15,~2012 (here onwards session~A and B,
respectively). A very strong, dispersed pulse with a
signal-to-noise ratio (S/N) more than 200, was discovered at
a DM of $\sim2.1$~pc~cm$^{-3}$ in session A. Both the
sessions~A and B were conducted in the day time, and suffered
heavily from RFI. Although prominent RFI contaminated spectral
channels as well as time samples were identified and excluded,
several low-level RFI intervals escaped our threshold based
RFI-identification algorithm. These remnant RFI intervals
reduced the sensitivity towards faint transient events.
Dynamic spectra of session~A and B were manually inspected
to identify the remnant potential RFI intervals.
In total, about 25\% of the time-samples were flagged
as RFI-contaminated. The flagged samples were excluded from any further
processing, including the repeated transient search.
The repeated search uncovered
several other dispersed pulses. A couple of bright pulses
discovered in session~A are shown in Figure~\ref{fig_dpulses}.
All the bursts discovered in session~A and B are listed in
Table~\ref{table_smry}.
\subsection{Dispersion Measures of the Bursts}
\begin{figure*}
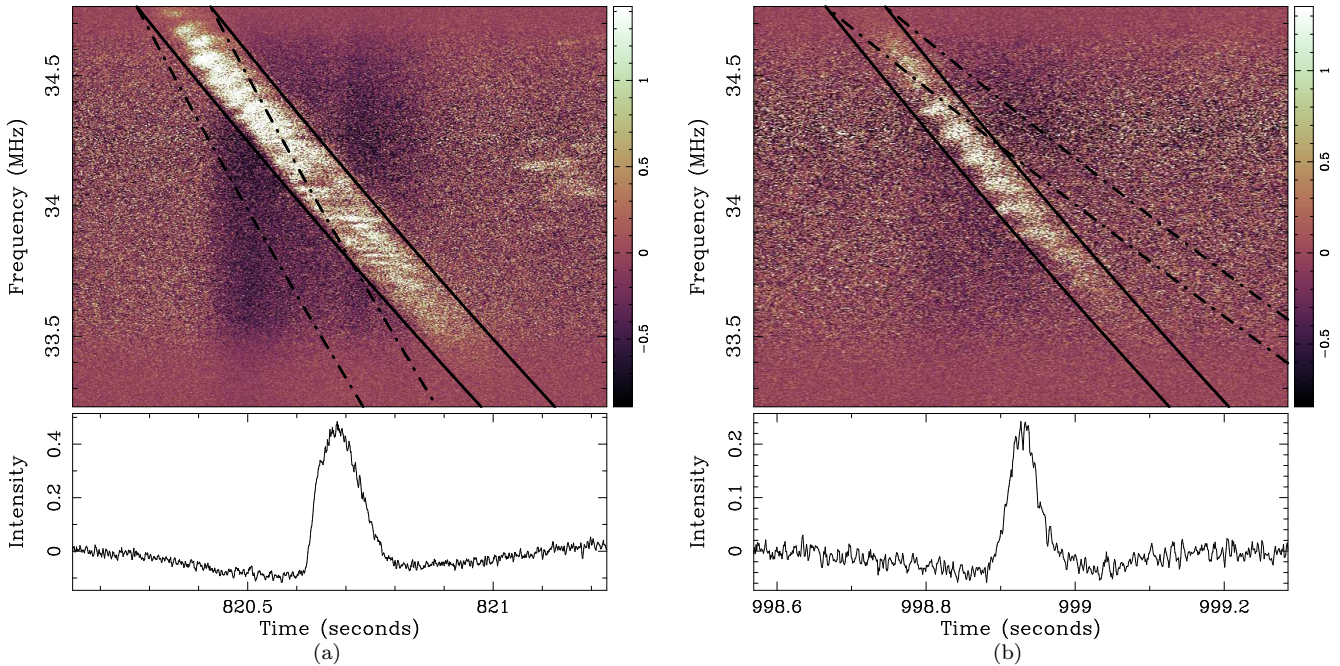

\centering
\subfigure[]{\includegraphics[width=0.47\textwidth,angle=-90]{fpa_0633p0632_dyn_spec_0820.69.ps}
\label{fig_dpulses1}}
\hspace*{2mm}
\subfigure[]{\includegraphics[width=0.47\textwidth,angle=-90]{fpc_0633p0632_dyn_spec_0998.93.ps}
\label{fig_dpulses2}}
 \caption{{\sl Two bright radio bursts discovered in session~A:}
In each of the subfigures (a) and (b), the upper panel shows a part
of the dynamic spectrum centered at the position of the burst.
The pairs of continuous slanted lines that bound the pulse, show the expected
quadratic behavior for the cold-plasma dispersion relation, corresponding
to the respective best-estimated DMs --- $2.16$~pc~cm$^{-3}$ and
$1.42$~pc~cm$^{-3}$ for the bursts in subfigures (a) and (b), respectively.
To demonstrate the difference in the two DM values, the dash-dot-dashed
lines in subfigure (a) shows the dispersion relation corresponding to the
DM of the burst in (b), and vice-versa. The lower panels show the total
intensity as a function of time measured from the start of the observation,
after dispersive delays are corrected using the corresponding best-estimated
DMs (for a reference frequency of 34~MHz). The intensity in both the panels
is in units of system temperature.}
\label{fig_dpulses}
\end{figure*}
Propagation of a radio signal through the ionized interstellar medium (ISM)
introduces frequency dependent delay in its arrival time. Considering the ISM
to be cold plasma comprised of free electrons, the delay ($\Delta t$) has a
power law dependence on frequency ($\nu$) and the dispersion measure (DM;
the integrated column density of free electrons along the line of sight):
$\Delta t \propto {\rm DM}\,\nu^{-2}$. As shown by the reference lines
indicating the expected quadratic delay as a function of frequency in
Figure~\ref{fig_dpulses}, the radio bursts closely follow the expected
cold plasma dispersion relationship. A quantitative fit of the observed
frequency dependence has not been possible due to presence of a mixed
variety of spectral features (some of which we use to deduce polarization
properties of the bursts as explained below). The estimated DMs of all the
radio bursts sample a range of 1.4--3.6~pc~cm$^{-3}$ (see Table~\ref{table_smry}).
The uncertainty in DM is measured by examining the significance of the
dedispersed pulse as a function of trial DM, and corresponds to a change
which degrades the peak amplitude of the pulse by $1\,\sigma$. The measured
uncertainties have been successfully cross-checked against the theoretical
predictions \citep[using equation~12 of][]{CM03}. The significant difference
in the DMs of different bursts is apparent in Figure~\ref{fig_dpulses},
wherein the quadratic dispersion relation curves corresponding to DMs
of two pulses are shown together.
%
\begin{deluxetable*}{lrcrcr}
\tabletypesize{\footnotesize}
\tablecolumns{6}
\tablewidth{0pt}
\tablecaption{Properties of all the radio bursts discovered in session~A and B.
The arrival times (pulse-centroids) are given with respect to the start
of observations, pulse-widths
are rounded-off to nearest multiple of 5, and pulse-energies are rounded-off to
nearest multiple of 100. The pulse-energies inside square brackets assume the
Geminga pulsar to be the source of radio bursts, while those outside the brackets
assume the source to be at the telescope pointing center. \emph{Absolute} values
of RM are given, since the sign can not be inferred.}
\tablehead{
           \colhead{Sr. No.}                                 &
           \colhead{Time of arrival}                         &
           \colhead{Pulse width}                             &
           \colhead{Pulse Energy}                            &
           \colhead{Dispersion Measure}                      &
           \colhead{Rotation Measure\tablenotemark{$\dagger$}} \\
           \colhead{}                        &
           \colhead{(seconds)}               &
           \colhead{(ms)}                    &
           \colhead{(Jy.ms)}                 &
           \colhead{(pc~cm$^{-3}$)}          &
           \colhead{(Rad.~m$^{-2}$)}         
          }
\startdata
\multicolumn{6}{l}{\textsl{\underline{Session~A (MJD at start of observation: 56121.247581019)}}}\vspace*{1mm}\\
 1 &  $  820.69\pm0.03$  & 110 & 403000~~\big[556900\big]  &  $2.16\pm0.07$  &       ---     \\
 2 &  $  823.53\pm0.02$  &  75 & 174000~~\big[240500\big]  &  $2.01\pm0.04$  &  $ 9.4\pm0.8$ \\
 3 &  $  824.43\pm0.03$  &  85 & 159700~~\big[220700\big]  &  $2.08\pm0.06$  &  $12.6\pm2.8$\tablenotemark{$\ddagger$} \\
 4 &  $  944.11\pm0.03$  & 140 &  81000~~\big[111900\big]  &  $2.62\pm0.13$  &       ---     \\
 5 &  $  998.93\pm0.02$  &  55 &  89600~~\big[123800\big]  &  $1.42\pm0.04$  &  $ 7.3\pm0.4$ \\
 6 &  $ 1000.41\pm0.02$  &  90 &  60300~~\big[ 83400\big]  &  $1.40\pm0.08$  &  $ 7.0\pm0.7$ \\
 7 &  $ 1157.89\pm0.03$  &  95 &  22200~~\big[ 30600\big]  &  $2.30\pm0.15$  &  $ 6.3\pm1.1$ \\
 8 &  $ 1169.80\pm0.03$  & 180 &  58600~~\big[ 81000\big]  &  $2.09\pm0.20$  &       ---     \\
 9 &  $ 1172.50\pm0.01$  &  50 &  21600~~\big[ 29800\big]  &  $1.45\pm0.11$  &       ---     \\
10 &  $ 1176.86\pm0.01$  &  35 &  26600~~\big[ 36800\big]  &  $1.42\pm0.10$  &  $10.6\pm1.3$ \\
11 &  $ 1177.08\pm0.01$  &  30 &  26400~~\big[ 36400\big]  &  $1.47\pm0.07$  &  $10.7\pm1.1$ \\
12 &  $ 1186.41\pm0.04$  & 170 &  18500~~\big[ 25500\big]  &  $2.59\pm0.40$  &       ---     \\
13 &  $ 1187.39\pm0.02$  & 195 &  28300~~\big[ 39000\big]  &  $1.74\pm0.27$  &       ---     \\
\multicolumn{6}{l}{\textsl{\underline{Session~B (MJD at start of observation: 56123.232627315)}}}\vspace*{1mm}\\
14 &  $  713.10\pm0.04$  & 130 &  57100~~\big[ 79000\big]  &  $3.62\pm0.17$  &       ---     \\
15 &  $ 1345.99\pm0.07$  & 270 &  91300~~\big[126200\big]  &  $2.91\pm0.17$  &       ---     \\
16 &  $ 1518.90\pm0.05$  & 150 & 106600~~\big[147300\big]  &  $3.41\pm0.11$  &       ---     \\
\enddata
\tablenotetext{$\dagger$}{Power in some of the bursts do not extend throughout
the full observed bandwidth of 1.5~MHz, and hence correspondingly limited
parts of the bandwidth were used in the RM-synthesis of these bursts. For
bursts 3, 6, 7, 10 and 11, \emph{effective} bandwidths used to estimate RM are
0.3, 1.0, 0.6, 0.5 and 0.6~MHz, respectively.}
\tablenotetext{$\ddagger$}{The Faraday spectrum of this burst is not well
resolved, and RM corresponding to the brightest component is provided here.}
\label{table_smry}
\end{deluxetable*}
%
\subsection{Deducing Linear Polarization Properties and Faraday Rotation from
Spectral Intensity Modulation}
%
Several of the radio bursts exhibit a variety of spectral structures,
sometimes at frequency scales as small as a few tens of kHz (see e.g.,
Figure~\ref{fig_dpulses1}). These spectral features may have been
contributed by, either individually or a combination of, the intrinsic
spectrum of the bursts and the propagation effects. The narrow bursts
shown in Figure~\ref{fig_dpulses} extend nearly throughout the observed
bandwidth. Although intrinsically narrow emission bandwidth is possible,
it is unlikely for an astronomical source to have narrow, systematic
spectral features within an intrinsically (relatively) wideband emission.
The relevant propagation effects include interstellar scintillation and the
Faraday rotation in the magneto-ionic medium. The scintillation bandwidth
towards the direction of our observation, estimated using the NE2001 model
\citep{CL02}, varies from a few kHz to a few tens of kHz for the DM range
of 1.4--3.6~pc~cm$^{-3}$, and might explain some of the observed spectral
features.
\par
Faraday rotation in the intervening magneto-ionic medium rotates the plane
of the linear polarization. {Rotation Measure (RM)} --- the line-of-sight
component of the intervening magnetic field weighted by the electron density
and integrated over the distance --- characterizes the amount of
frequency-dependent rotation: $\Delta\theta = {\rm RM} \lambda^2$, where
$\lambda$ is the wavelength of observation. Note that the Gauribidanur radio
telescope is receptive to only a single linear polarization (in east-west
direction). For such a telescope, the differential Faraday rotation of the
linearly polarized component of the signal results in spectral intensity
modulation within the observation bandwidth \citep[see, e.g.,][]{RD99}.
The observed spectral intensity modulation can be exploited to estimate RM,
and study the linear polarization properties (i.e., degree of polarization and
the polarization position angle) of the signal. For this purpose, we have
used an approach similar to the ``auto-correlation domain approach'' described
by \citet{RD99}. To estimate RM, we take the radio frequency spectrum at each
of the time samples within the pulse, transform the frequency values
corresponding to different spectral channels to square of their respective
wavelengths ($\lambda$), and take a discrete Fourier transform of the
spectrum in the $\lambda^2$-domain. The Fourier conjugate of $\lambda^2$
is directly proportional to RM, and hence the above discrete Fourier
transform provides linearly polarized power as a function of RM. In this
`Faraday spectrum', the spectral modulation due to the Faraday rotation would
appear as a narrow feature convolved with the Fourier transform of the
filter response (i.e., bandshape). Note that in the absence of prior knowledge
of the polarization position angle (PPA) sweep across the pulse, the above
process needs to be performed separately on spectra corresponding to the
individual samples, amounting to a \emph{time-resolved RM synthesis}.
However, in this method, the RM estimation is limited by S/N of individual
samples. A more sensitive estimate of RM is obtained by
incoherently adding the spectral modulation power in various samples, i.e.,
by taking  a weighted or simple average of the Faraday amplitude spectra
across the pulse. After estimating the RM, fractional linear polarization
and PPA can be estimated across the pulse by using the amplitude and phase
of the corresponding feature in the Faraday spectra.

\par
The above method of deducing RM and linear polarization properties was
successfully tested using simulated data as well as known pulsars observed
using Gauribidanur and Ooty radio telescopes (at 34 and 326.5~MHz,
respectively).
At 34~MHz, RM as well as linear polarization properties of B0950$+$08
were deduced using filterbank data (with 512 channels across 1.53~MHz bandwidth
and a time resolution of 1.66~ms) from a 30~minutes long observation. The deduced
RM of $2.8\pm0.4$~rad$\,$m$^{-2}$ and percentage linear polarization of about
60\% are consistent with the earlier published values of 2.15~rad$\,$m$^{-2}$
and 73.9\%, respectively \citep[at 150~MHz;][]{Noutsos15}. At 326.5~MHz, an
approximately 5~minutes long observation of the bright pulsar B0740$-$28 using
the new software backend \citep{Naidu15} of the Ooty radio telescope was utilized.
Using the data in filterbank format with time and frequency resolutions of about
1~ms and 15.6~kHz, respectively, the deduced RM ($155\pm5$~rad$\,$m$^{-2}$) and
percentage linear polarization (about 92\%) of B0740$-$28 are consistent with
the earlier reported values at 435~MHz \citep[150.4~rad$\,$m$^{-2}$ and 83\%,
respectively;][]{GL98,MHQ98}. The deduced PPA sweeps of the two pulsars were
also found to be consistent with those reported at nearby frequencies.
\par
\begin{figure*}
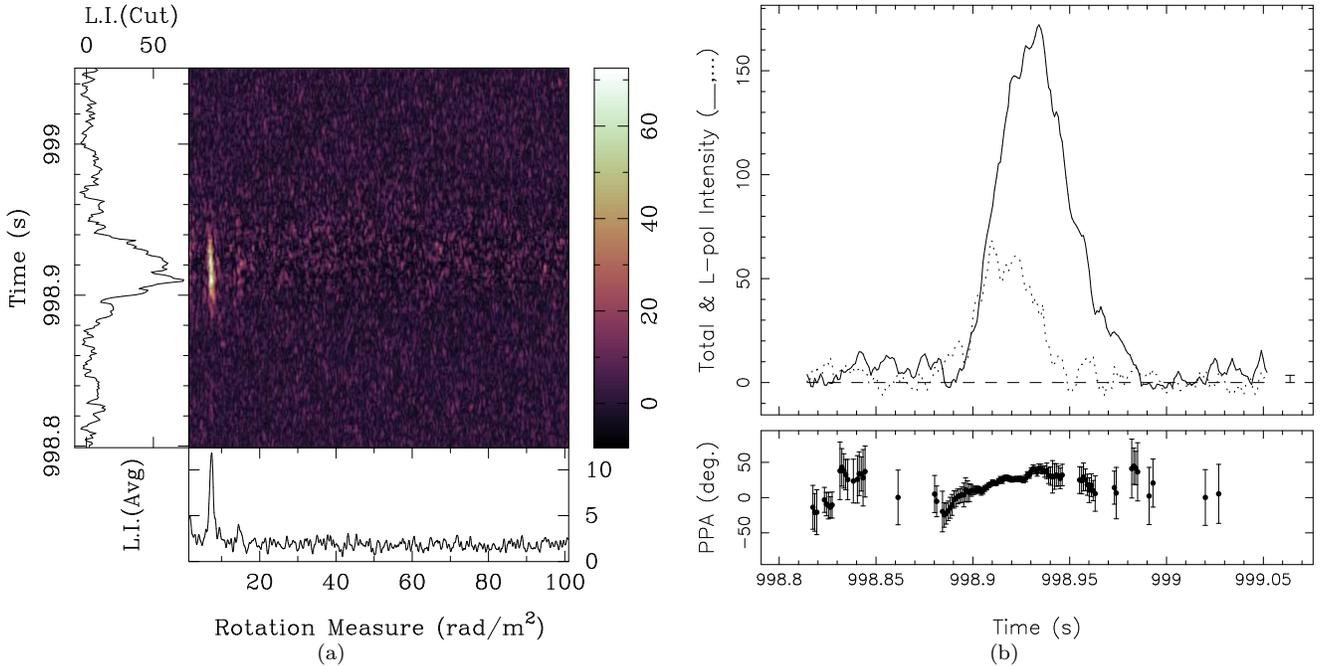

\centering
\subfigure[]{\includegraphics[width=0.47\textwidth,angle=-90]{fqa_rm_extract_0998.93.ps}
\label{fig_rm1}}
\hspace*{2mm}
\subfigure[]{\includegraphics[width=0.47\textwidth,angle=-90]{fqb_0633p0632_GBD_20120713_polProf_0998.927.ps}
\label{fig_rm2}}
 \caption{{\sl Rotation measure synthesis and linear polarization properties
of the burst shown in Figure~\ref{fig_dpulses2}:} The central panel in
subfigure~(a) shows Faraday amplitude spectra (linearly polarized intensity
as a function of RM) at different times around the burst location. The bottom
and left panels show the average Faraday amplitude spectrum and a vertical
cut across the maximum in the central panel, respectively. The upper panel
in subfigure~(b) shows the total (continuous line; in arbitrary units) and
the linearly
polarized intensity (dotted line; corresponding to RM=$7.3\pm0.4$~rad$\,$m$^{-2}$)
profiles of the burst. The lower panel shows the PPA behavior across the burst.}
\label{fig_rm}
\end{figure*}

In Figure~\ref{fig_rm1}, a {time-resolved RM synthesis}
performed on the burst in Figure~\ref{fig_dpulses2} is shown.
Using the average Faraday amplitude spectrum,
RM for this pulse is estimated to be $7.3\pm0.4$~rad$\,$m$^{-2}$. Having estimated
the RM, fractional linear polarization and PPA can be estimated across the
pulse by using the amplitude and phase of the corresponding feature in the
Faraday spectra. The linear polarization properties of the burst obtained
in this way are shown in Figure~\ref{fig_rm2}.
\par
\begin{figure*}
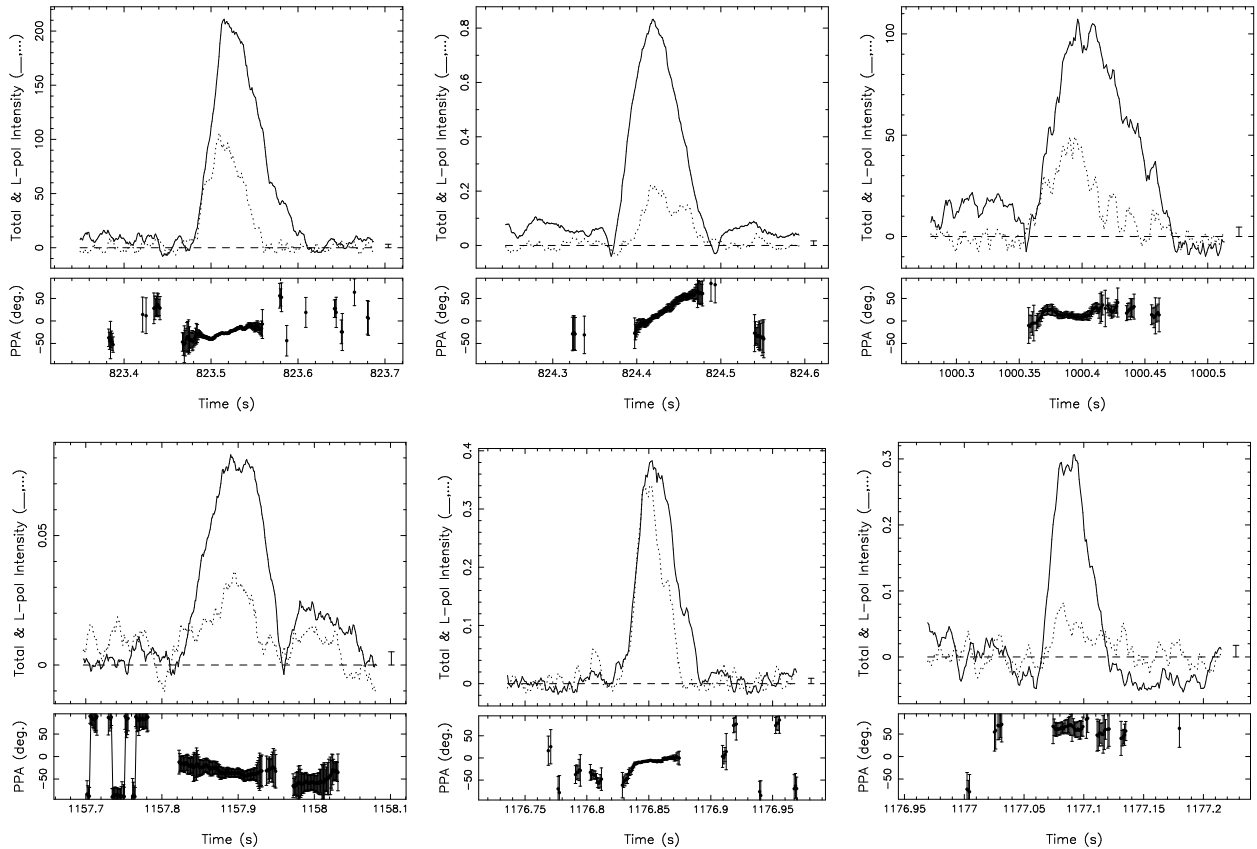

\centering
\subfigure{\includegraphics[width=0.3\textwidth,angle=-90]{fra_0633p0632_GBD_20120713_polProf_0823.530.ps}}
\hspace*{2mm}
\subfigure{\includegraphics[width=0.3\textwidth,angle=-90]{frb_0633p0632_GBD_20120713_polProf_0824.422.ps}}
\hspace*{2mm}
\subfigure{\includegraphics[width=0.3\textwidth,angle=-90]{frc_0633p0632_GBD_20120713_polProf_1000.418.ps}}
\vspace*{2mm}
\subfigure{\includegraphics[width=0.3\textwidth,angle=-90]{frd_0633p0632_GBD_20120713_polProf_1157.893.ps}}
\hspace*{2mm}
\subfigure{\includegraphics[width=0.3\textwidth,angle=-90]{fre_0633p0632_GBD_20120713_polProf_1176.869.ps}}
\hspace*{2mm}
\subfigure{\includegraphics[width=0.3\textwidth,angle=-90]{frf_0633p0632_GBD_20120713_polProf_1177.091.ps}}
\caption{{\sl Linear polarization profiles of radio bursts:} Linear polarization
properties (continuous line: total intensity in arbitrary units, dotted line:
linear polarization
intensity, and PPA in the lower panels) of six radio bursts corresponding to their
estimated RMs (see Table~\ref{table_smry}) are shown.}
\label{fig_polprofs}
\end{figure*}

The pattern of spectral features in the observed dynamic spectra of various bursts
is likely to be a superposition of the spectral intensity modulation due to Faraday
rotation and random modulations due to interstellar scintillation on the scale
of scintillation bandwidth. Consequently, in the Faraday spectrum, the feature
due to Faraday rotation is also convolved with the Fourier transform of the
modulations due to scintillation. The random modulation are smeared out if the
signal can be averaged over spans much longer than the scintillation time. In
our case of RM synthesis using single pulses, this effect is more relevant and
generally increases the uncertainty in RM estimation, and in some cases even
limits the estimation of RM (e.g., estimation for the burst shown in
Figure~\ref{fig_dpulses1} was not possible, although it seems consistent with
a RM of about 10~rad~m$^{-2}$). As mentioned earlier, estimation
of RM is also limited by the S/N of individual samples. Despite these limitations,
RM could be estimated unambiguously for 7 bursts discovered in session A, and
these estimates\footnote{The RM estimates are obtained from a (S/N)$^2$-weighted
average of Faraday spectra across the pulses. However, the uncertainties
mentioned in these RM estimates are those obtained from individual Faraday
spectra. We mention these larger uncertainties to take in to account any effect
of convolution of filter response or scintillation feature in the Fourier domain.
The actual uncertainties could be better (i.e., smaller) by 50--80\% of those
 mentioned here.} are listed in Table~\ref{table_smry}. The linear polarization
properties of these 7 bursts are shown in Figures~\ref{fig_rm2} and
\ref{fig_polprofs}.
%
%
%
\subsection{Search for Transients toward Control Sources and Blank Sky}
The control source observations were analyzed to detect periodic signals
from the known pulsars and to search for any transient events. Average
periodic signals as well as infrequent bright pulses from the control
pulsars B0834+06 and B1919+21 were regularly detected at their corresponding
DMs. On the day corresponding to session-A observations, both the control pulsars
were observed. B1919+21 was observed about 11~hours before the session-A,
while B0834+06 was observed about 2~hours after the session. Towards B1919+21,
no genuine single pulse candidate was found above $8\,\sigma$ significance.
B0834+06 observation was carried out in the afternoon (one of the peak
times for RFI at the observatory), and was severely affected by RFI
(much more than the session-A). Although such severe RFI made the search
for single pulses nearly impossible, a search and manual inspection
involving thorough investigations of the dynamic-spectra did not result
in detection of any genuine transient candidate in the RFI-free parts of
data. Several other gamma-ray pulsars (J1732$-$3131, J1809$-$2332,
J1836$+$5925, J2021$+$4026, J2055$+$2539, J2139$+$4716) were also observed
on the same as well as the previous day of session-A observations. No
significant dispersed pulse was found in these observations. Average
periodic signal from both the pulsars (despite the severe RFI conditions
during B0834+06 observation) could be detected at $\sim 10\,\sigma$ level,
providing confidence in the system-health.
\par
In the absence
of any known radio pulsars or detection of signals from the target sources,
our pointings towards all the gamma-ray pulsars effectively amount to
observing the blank sky. Search for transient events in these blank sky
observations (nearly 250 individual sessions of 30~minutes each) also did not
result in detection of any significant dispersed pulse \citep{MA14}.
%
%
\section{Fast Radio Transients: Possible Source and Emission Mechanism}
\subsection{Source of the radio bursts}
In the previous section, we presented discovery and properties of several
fast radio transients with pulse widths in the range 30--270~ms. The radio
bursts exhibit two important properties which are likely to be associated
with their propagation through the ISM. Their times of arrival across the
observation bandwidth closely follow the behavior expected from the
cold-plasma dispersion relation, and several of the bursts also exhibit spectral
intensity modulations that are likely to be due to differential Faraday
rotation of the plane of linear polarization within the observing bandwidth.
\par
Sometimes, the {swept-frequency RFI} can mimic the dispersive signature
of astronomically originated signals across the bandwidth. However, the
times of arrival of a swept-frequency RFI are generally linearly proportional
to the frequency, as against the $\nu^{-2}$ dependence in the dispersion
relationship (Section 3). As a part of further critical assessment,
data corresponding to the
brightest radio burst (shown in Figure~\ref{fig_dpulses1}) were dedispersed
using linear delay gradients ($\Delta t \propto \nu$) spanning adequate
delay ranges. The significance of detection in this case was found to be
much lower compared to that obtained using the dispersion relation
($\Delta t \propto {\rm DM}\,\nu^{-2}$), suggesting
the signal to be of astronomical origin.
Furthermore, several of the radio bursts also exhibit spectral modulation
corresponding to the Faraday rotation in the intervening magneto-ionic
medium. It would require a conspiracy for the swept-frequency RFI to mimic
the dispersive as well as the Faraday rotation signature of astronomical
signals. Note that the maximum ionospheric contribution towards RM, during
the observing session~A, is estimated\footnote{The Earth's ionosphere also
contributes to the
observed RM, and its contribution depends on the total electron content and
the Earth's magnetic field component along the line of sight. To estimate
the amount of ionospheric Faraday rotation for the specific epoch, geographic
location, and line-of-sight corresponding to our observations, we used the
software {\sf ionFR} \citep[][\url{http://sourceforge.net/projects/ionfr/}]{ionfr13}.
{\sf ionFR} uses the International Geomagnetic Reference Field and a number
of publicly available, GPS-derived total electron content maps. Since the
telescope beam-width is very large ($\sim$25\mdeg) in declination, we examined
several lines of sight within the beamwidth, and the maximum ionospheric RM
was found to be 2~rad~m$^{-2}$.} to be only 2~rad~m$^{-2}$ --- much smaller
than RM of various bursts (6--10~rad~m$^{-2}$). Hence, the deduced
RM values also indicate the source of radio bursts to be much beyond the Earth's
ionosphere (i.e., at astronomical distances).
Several of the radio bursts also show indications of an exponential tail in
their pulse shapes (e.g., the bursts shown in Figure~\ref{fig_dpulses}, that
in top-left subplot of Figure~\ref{fig_polprofs}, and a couple of bursts not
shown) --- a characteristic often associated with interstellar scattering of
pulsed signals\footnote{Large pulse-widths, spectral structures due to
scintillation and Faraday rotation, and small amount of apparent scattering
(as expected for the correspondingly small DMs), make it inadequate to test
the power-law dependence of the scatter-broadening across the bandwidth.}.
Non-detection of any transient events towards several other sky
directions (see Section 3) also strongly supports the inference that the
bursts could not have been a result of some man-made or system-originated signal.
Hence, it is nearly certain that these radio bursts have their origin in an
extraterrestrial source.
\par
Although the estimated DMs of various radio bursts are not consistent
with a single DM value, and lie in the range 1.4--3.6~pc~cm$^{-3}$
(Table~\ref{table_smry}), it is unlikely that the bursts corresponding
to different DMs originated from different radio sources that were
coincidently active only within the short spans of 2 observing sessions
(A \& B; out of 130 sessions). The reason for variation in DM is more
likely to be associated with the source or the intervening medium.
With this argument, the following discussion proceeds with the assumption
that all the radio bursts have been originated from a single astronomical
source.
\subsubsection{Could the bursts have originated from Sun ?}
Type II/III solar bursts \citep[see, for example,][]{solar85} can sometime
mimic the dispersive signature of distant astronomical pulsed signals.
The spectral slopes or the frequency drift rate of these bursts depend on
the plasma density and could vary from one burst to the other. However,
widths of type III bursts are typically of the order of 1~s or more, and
those of type II bursts are of the order of several minutes, i.e., much
more than the widths of radio bursts under consideration. Furthermore, so
far there is no evidence for solar radio bursts to have any significant
linear polarization. Also, during the observing sessions A and B, Sun's
position was about 15\mdeg~away from the pointing RA (beamwidth in RA is
$\approx 0.35$\mdeg). Hence, bursts from Sun could have been detected only
from a very far side-lobe, implying a huge attenuation in measured burst's power.
These aspects, especially the observed narrow widths and significant linear
polarization, make it unlikely that these bursts could have been originated
from the Sun. To further explore any minor possibility of these bursts being
associated with the Sun, we used the archived solar monitoring data from
IZMIRAN's Solar Radio Laboratory\footnote{\url{http://www.izmiran.ru/stp/lars/}}
\citep[][]{Gorgutsa01} with a time resolution\footnote{Originally, these data
were recorded with a finer resolution of 0.04~s. However, before archiving,
the time resolution was coarsened to 1~s by taking an average over consecutive
25 samples.} of 1~second. A part of these data corresponding to a bandwidth
of about 1.5~MHz centered at 34~MHz were examined, and no event was found
around the epochs of the radio bursts discovered in session A and B.
An examination of archival solar monitoring data from the Learmonth and
Culgoora observatories, albeit with a coarse resolution of 3~s, also did
not indicate any event corresponding to the epochs of radio bursts under
consideration. Hence, it is conclusive that these radio bursts have not
originated from the Sun.
\subsubsection{A pulsar as a possible source of radio bursts}
Pulsars are known to emit bright radio pulses with high degree of linear
polarization. Some pulsars also emit in the form of extremely bright and
narrow pulses called {giant-pulses}. However, the observed short timescale
variation in DM is unprecedented. Arrival times of individual bursts
depend on their DM, and a typical uncertainty of 0.1~pc~cm$^{-3}$ in DM
would imply a corresponding uncertainty of about 360~ms in arrival times.
Hence, search for a periodic signal, as expected from a pulsar, across the
observing session would not be adequate.
Nevertheless, we performed blind periodicity searches around the
DMs suggested by the bursts. Although no significant periodic signal was
found in these blind searches, we can still examine if the
bursts have originated from a known pulsar within the beam.
%
\begin{figure}
\centering
\subfigure[]{\includegraphics[width=0.25\textwidth,angle=-90]{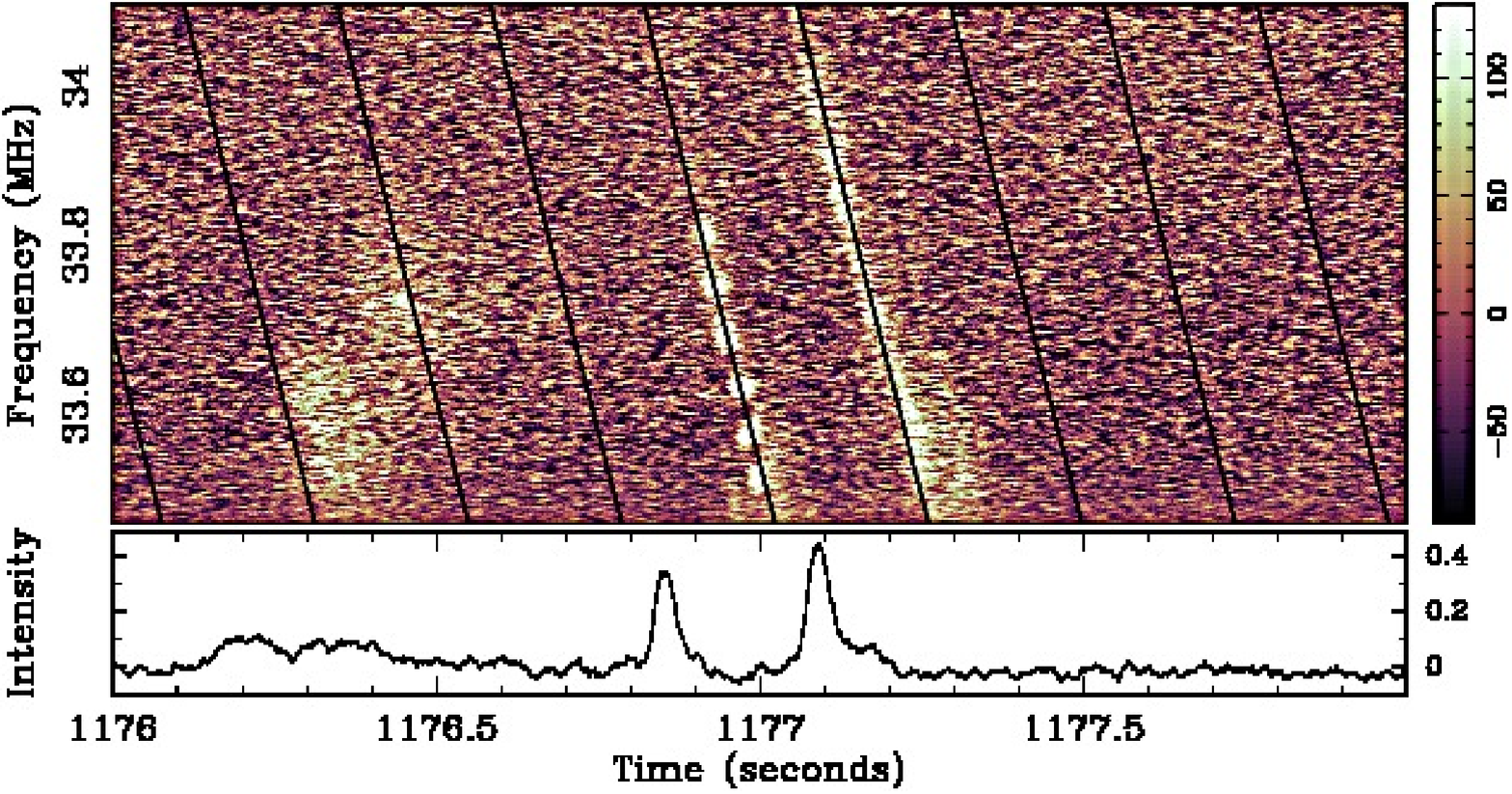}
\label{fig_dspecs1}}
\vspace*{2mm}
\subfigure[]{\includegraphics[width=0.25\textwidth,angle=-90]{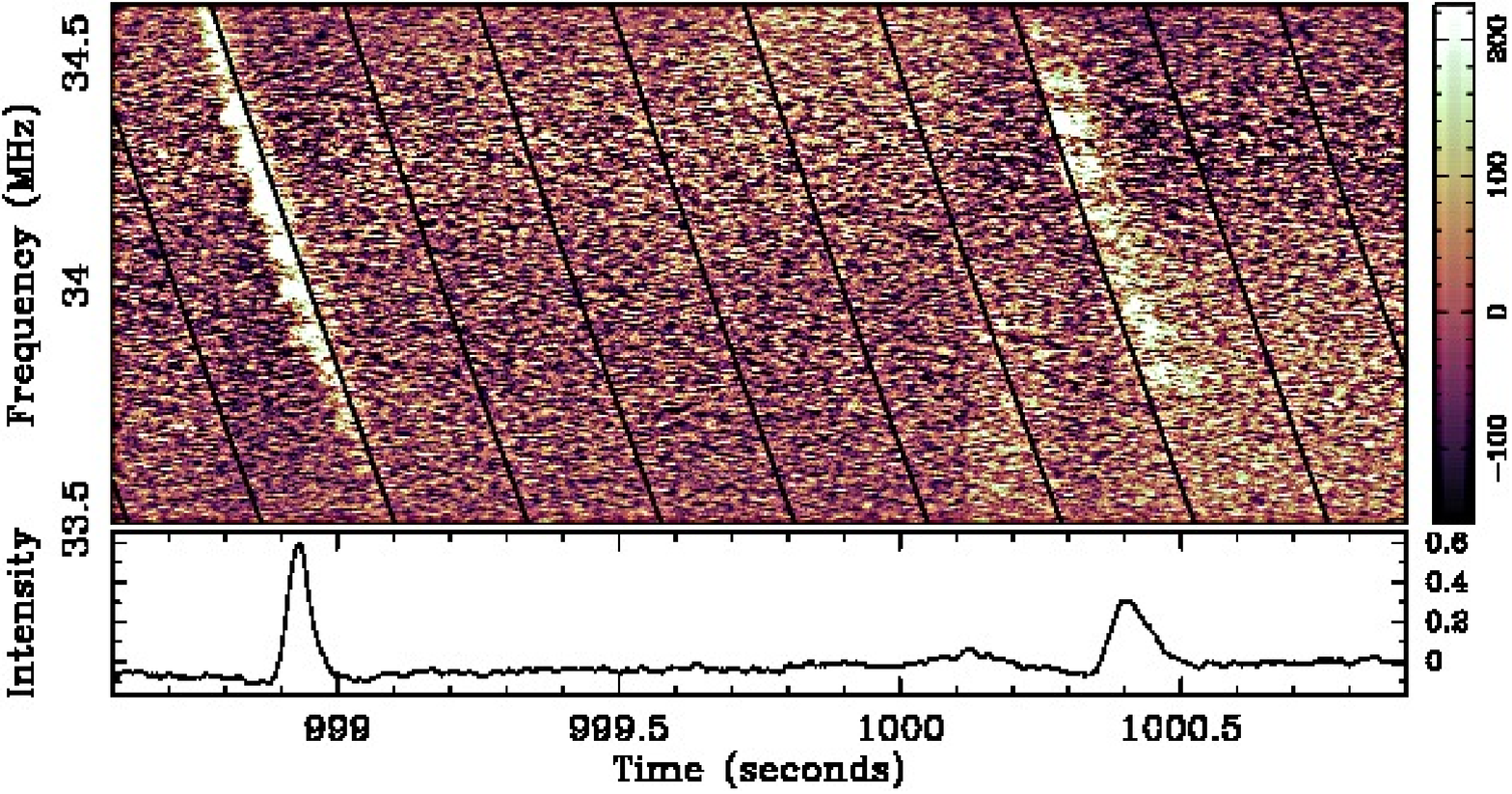}
\label{fig_dspecs2}}
\vspace*{2mm}
\subfigure[]{\includegraphics[width=0.25\textwidth,angle=-90]{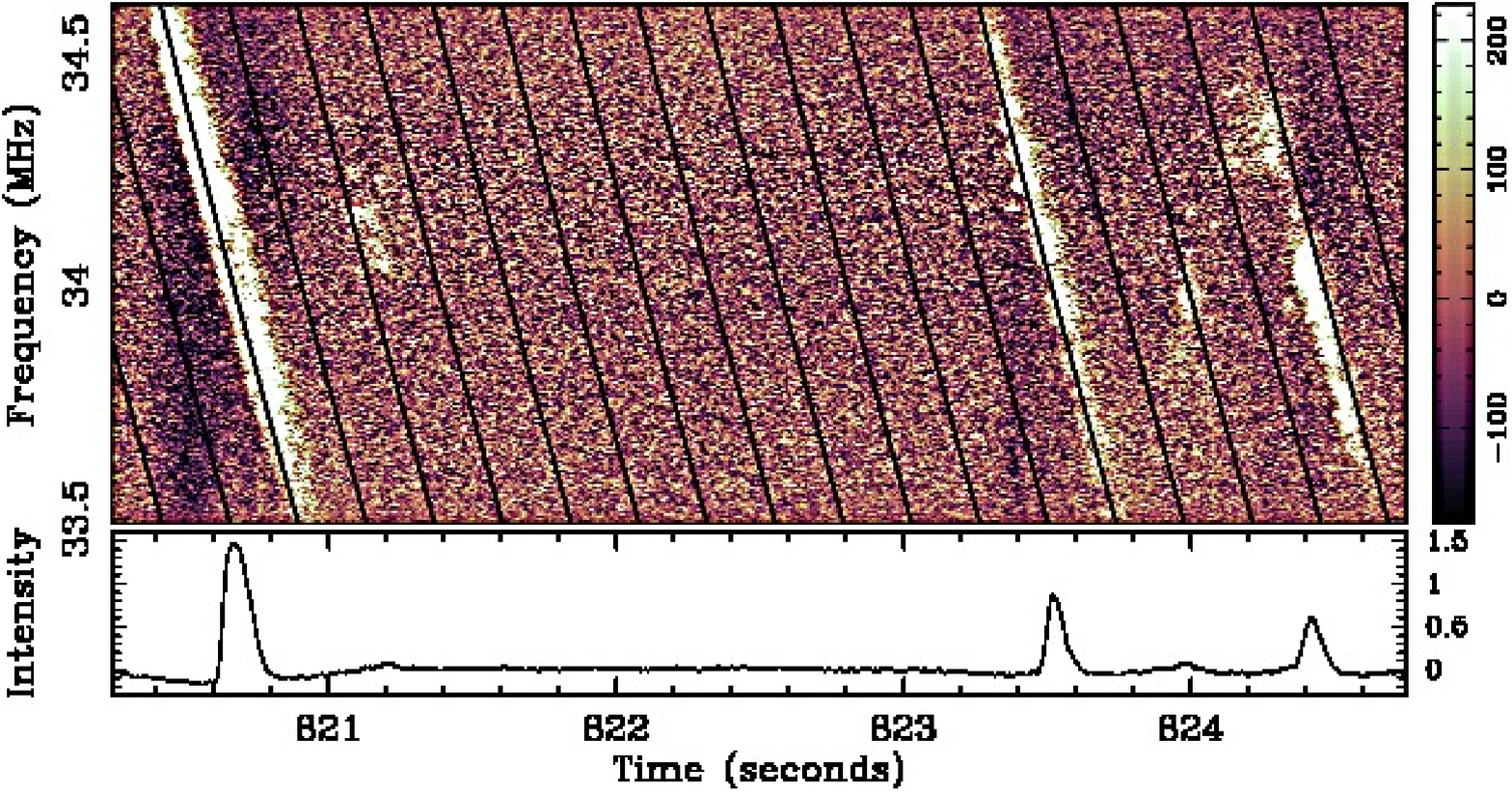}
\label{fig_dspecs3}}
\caption{{\sl Spectrograms around groups of radio bursts:}
{In each of the subfigures, the upper panel shows spectrogram around the
selected group of bursts. The slanted thick black lines show the expected
periodic times of arrival of radio pulses from Geminga pulsar when
dispersed with DMs corresponding to that of bursts in subfigures (a),
(b) and (c), i.e., 1.45, 2.05 and 1.40~pc~cm$^{-3}$, respectively.
The lower panels show the time series after the dispersion delays are corrected
for a reference frequency of 34~MHz and using the above mentioned DMs.
The spectrogram in (a) is limited to about 0.6~MHz (since the corresponding
group of bursts do not extend beyond this bandwidth), while those in (b)
and (c) are shown for parts of the bandwidth where the filter-response is
nearly flat ($\sim$1.1~MHz).}}
\label{fig_dspecs}
\end{figure}
\begin{figure*}
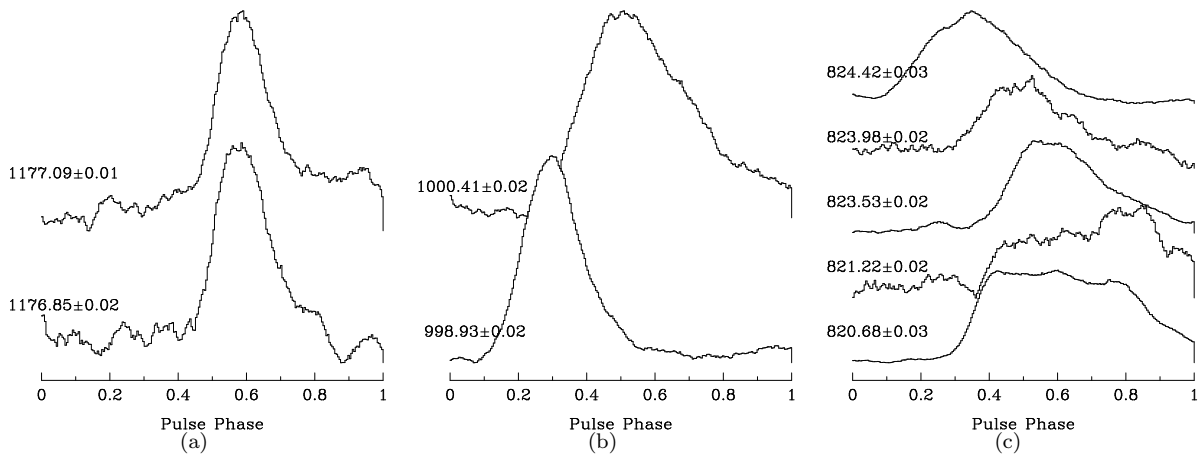

\centering
\subfigure[]{\includegraphics[width=0.31\textwidth,angle=-90]{ftc_pstack_1172.ps}}
\hspace*{2mm}
\subfigure[]{\includegraphics[width=0.31\textwidth,angle=-90]{ftb_pstack_0998.6.ps}}
\hspace*{2mm}
\subfigure[]{\includegraphics[width=0.31\textwidth,angle=-90]{fta_pstack_820.10_33.9_34.2.ps}}
\caption{{\sl Distribution of radio bursts in Geminga's rotation phase:}
The panels (a), (b), and (c) show the bursts displayed in the spectrograms
of Figure~\ref{fig_dspecs1}, \ref{fig_dspecs2}, \& \ref{fig_dspecs3}, respectively,
when dedispersed, and stacked according to their distribution in rotation phase
of the pulsar. A small bandwidth of only 0.3~MHz (33.9--34.2~MHz) has been
used for the bursts shown in panel (c), to avoid S/N-attenuation of the
narrow band bursts. For each of the bursts, its arrival time (pulse-centroid
in seconds; from start of observation) is shown on the left side.}
\label{fig_stacks}
\end{figure*}
\par
The observations were originally targeted at the gamma-ray pulsars
J0633+1746 and J0633+0632. To explore if the radio bursts have originated
from one of these two pulsars, we note that there are groups of radio bursts which
were detected within a few seconds duration and have DM as well as RM values
consistent with each other. Three such groups comprise of burst numbers 1--3,
5--6, and {10--11} (hereafter, group A1, A2, and A3, respectively) in
Table~\ref{table_smry}. Within each of these groups, we can examine if the
arrival times of the bursts are consistent with the rotation
period\footnote{The expected rotation period at the observation epoch is
obtained by using the timing model provided by the LAT team at \texttt{
https://confluence.slac.stanford.edu/display/GLAMCOG/LAT+Gamma
-ray+Pulsar+Timing+Models} and the pulsar timing software
\textsc{Tempo2} (\url{http://www.atnf.csiro.au/research/pulsar/tempo2/}).}
of either of the two pulsars. Figure~\ref{fig_dspecs} shows spectrograms
(smoothed by a boxcar of width $\sim$15~ms)
around the above three groups of bursts, overlaid with expected arrival times
of periodic pulses from Geminga pulsar (J0633+1746; rotation period $\sim 0.237$~s).
The zero-phase of the expected periodic arrival times is different
in the three spectrograms and, is chosen manually (since uncertainty in DM
is inadequate for setting a common zero-phase).
Figure~\ref{fig_dspecs1}
clearly suggests that the group~A3 bursts (i.e., bursts 10 \& 11) are
consistent with being two consecutive radio pulses from Geminga pulsar,
separated by just one rotation of the pulsar. Other 5 bursts in the remaining
two groups (Figure~\ref{fig_dspecs2} and \ref{fig_dspecs3}) are also consistent
with them originated from Geminga pulsar and dispersed at the respective
DMs of the burst-groups. Moreover, Figure~\ref{fig_dspecs3} also shows
additional, narrow-band signals at around 821.1~s and 824~s, which give rise
to only hints of their presence in the time series. The arrival times of these
{narrow-band bursts}, along with those of the other three strong and wider-band
bursts, are also consistent with those expected from Geminga
pulsar\footnote{Figure~\ref{fig_dspecs1} also shows narrow-band signals
between 1176.25--1176.5~s. Given their broad widths, comparing their arrival
times with those expected from Geminga pulsar would not be adequate.}.
Arrival times of the two bursts
in Figure~\ref{fig_dspecs2} seem to be little offset from those expected from
the pulsar, but still consistent within a phase-window of $\sim$20--25\% of the
pulsar's rotation period. {Hence, arrival times of all the radio bursts within
the three groups (5, 2 and 2 bursts in group A1, A2 and A3, respectively)
strongly suggest that these bursts have originated from Geminga pulsar.}
\par
An examination of the distribution of dedispersed radio bursts in rotation
phase of Geminga pulsar, separately for three burst-groups, gives a further
clearer picture. As evident from Figure~\ref{fig_stacks}, first 4 bursts in
group~A1 (shown in panel (c)) and both the bursts in group~A3 (shown in
panel (a)) strongly suggest their origin to be Geminga pulsar. The pulse-profiles
shown in Figure~\ref{fig_stacks} are smoothed by a nearly 15~ms wide boxcar.
The fourth burst (from bottom) in Figure~\ref{fig_stacks}(c) appears to be
offset from the first three bursts by about 20~ms (nearly 10\% of the pulsar
period), however, given the effective resolution of about 15~ms and the
uncertainty in pulse-centroid, this offset is not significant.
Fifth burst in group~A1 and the two bursts in group~A2 are indeed slightly
offset, however the observed offset of about 20\% of the pulsar's rotation
period is still comparable with the widths of the bursts.
\citet{MM00} have shown histograms of widths and arrival-phases
of the average main and inter-pulse of Geminga. Their pulse-width ranges
(about 20--140 ms and 0--90 ms, for main and inter-pulse, respectively),
as well as the {average} phase-jitter range (about 0.2 of the period),
are consistent with the respective parameters of the bursts presented here.
\par
A similar examination of the spectrograms using the expected period of the
gamma-ray pulsar J0633+0632 indicates that arrival times of only two pairs of
bursts --- 2,3 and 5,6 --- seem to be consistent with the expected periodicity.
However, note that the expected periods of Geminga and J0633+0632 at the epoch
of our observation are about 0.2371 and 0.2974~s, respectively. Since the
typical pulse-widths of the bursts under consideration are of the order of
50~ms or more, the two periods can have several common multiples where the
arrival times of the bursts would appear to be consistent with both the
periods. Indeed the above two pairs of bursts suffer from this ambiguity.
However, the ambiguity is decisively resolved by the group A3 bursts, wherein
the difference between the arrival times ($220\pm20$~ms) clearly suggests
the period to be consistent with that of Geminga pulsar.
\par
Rotation periods of RRATs are generally derived by computing the largest
common divisor of the differences between the burst arrival times at a given
epoch \citep{McLaughlin06}. This approach can also be applied to estimate any
periodicity of the detected bursts independently. However, to ensure that the
differences in arrival times are intrinsic to the source and not due to any
variation in DM, the bursts only within the individual groups A1, A2 and A3
can be used for this purpose. Hence, we can use 4 independent arrival time
differences --- 2 differences from the arrival times of 3 wide-band
bursts in group A1, and 1 each from the pairs of bursts in groups A2 and A3.
A blind estimate of the period using these arrival time differences is
$232\pm8$~ms, which is consistent with the rotation period of Geminga and
further supports the above inference that the bursts have originated from
Geminga pulsar.
\par
To assess the chance probability of various bursts aligning with
Geminga's rotation phase, we performed Monte~Carlo (MC) simulations for the
three burst-groups. For group A3, it is significant that the two bursts
appear to be consecutive pulses from the pulsar.
To estimate the time-offset between the two bursts, brighter of the
two pulses is used as a template and cross-correlated with the timeseries
shown in Figure~\ref{fig_dspecs1}. From the cross-correlation function,
the delay offset between the two bursts is estimated to be in the range
223--242~ms. An
individual realization of MC simulation for this group involves generating
two arrival times randomly and uniformly distributed within the observation
duration of 1203~seconds. Using 10~million ($10^7$) such independent realizations,
the chance probability of the two bursts occurring with the above mentioned
delay is estimated to be $3.3\times10^{-5}$. In other words, the two bursts
are consistent with being two consecutive pulses from Geminga pulsar at a
confidence level of 99.997\%.
\par
For groups A1 and A2, the individual realizations of MC simulations
consist of generating 5 and 2 bursts randomly distributed within time-extents
of 5 and 2~seconds (as suggested by the durations within which these bursts
have been observed; see Figure~\ref{fig_dspecs}), respectively. Using 10~million
independent realizations, the probability that the 5 bursts in group A1 could
have aligned with each other within a phase-offset of $\pm0.3$ just by chance,
is found to be 0.00922. Similarly, the chance probability of one burst in group
A2 to be aligned with the other within a phase-offset of $\pm0.4$ is estimated
to be 0.42851.
Note that the above phase ranges sufficiently cover the
phase-offsets\footnote{A correlation analysis similar to that performed for
group~A3 bursts, by using the brightest burst in the individual groups as
a template and cross-correlating with the respective timeseries, was also
performed for groups A1 and A2. The phase-offsets between various bursts
suggested by this analysis are consistent within the phase-offset ranges we
have used.} between various bursts within the individual groups A1 and A2,
as apparent from Figure~\ref{fig_stacks}. Moreover, by using the phase-offset
ranges that extend on both sides of 0 (corresponding to perfect alignment),
we have taken a liberal approach, i.e., both the possible signs of the
relative offsets are allowed in the simulations.
From the above chance probabilities, the confidence levels at which the bursts
in groups A1 and A2 could be considered from Geminga pulsar, are 99.078\% and
57.145\%, respectively.
The low confidence level for group~A2 bursts to be associated with
the pulsar has resulted due to (1) presence of only two pulses in the group
which are also significantly offset from each other, and (2) our liberal
approach in choosing the corresponding phase-offset range.
\par
The net chance probability that the bursts in groups A1, A2 and A3
could have had the alignment with Geminga's rotation phase (within even
the large offsets stated above), is only $1.3\times10^{-7}$
($0.00922\times0.42851\times3.3\times10^{-5}$). Hence, it is nearly
certain, {i.e., at a confidence level of 99.999987\%}, that the bursts
have originated from the gamma-ray pulsar Geminga.
\par
We also note that the distance to the source estimated using NE2001 model
\citep{CL02} for the range of DMs sampled by various bursts in the two
sessions is in the range 100--320~pc, which is consistent with the
parallax distance to Geminga \citep[250$^{+120}_{-62}$~pc;][]{Faherty07}.
\par
Deep searches for persistent periodic signals from Geminga and
J0633+0632 using all the observing sessions also improve the previous upper
limits on respective flux densities \citep{MA14} by a factor of about 1.5.
The updated upper limits obtained by combining all the data are 25 and 19~mJy
for Geminga and J0633+0632, respectively.
\subsection{Possible emission mechanism of the radio bursts}
The energies of the radio bursts are extremely high, and pulse-energies
of this order have been observed to be emitted only from a handful of pulsars,
in the form of {giant pulses}.
The pulse energies of several radio bursts, including the brightest one,
are comparable with those of the giant-pulses observed from Crab
pulsar at decametre wavelengths \citep[at 23~MHz;][]{Popov06}. The comparable
pulse energies suggest the radio bursts to be giant pulses from Geminga
pulsar. The intrinsic widths of giant pulses have been observed to be very
narrow --- sometimes as small as a few nanoseconds \citep[][]{Hankins03}.
Relatively larger widths of the bursts (30--270~ms) might have
resulted intrinsically, or might have been caused by the same mechanism which
is responsible for the short timescale variation in their DM.
The giant pulse intensity and energy distributions exhibit power-law
statistics \citep[e.g.,][]{AG72}, in contrast to those of the regular pulses
which generally follow a normal distribution \citep[e.g.,][]{HW74}. Given the
statistically small number of bursts discovered in session~A and B, pulse
energy distribution is fitted equally well by power-law and normal
distributions. Overall, the observed extreme pulse-energies might suggest
the bursts to be radio giant pulses from Geminga pulsar.
\subsection{Cause of variation in DM}
The variations in DM of the bursts in session~A take
place at timescales\footnote{Variations at shorter timescales are also
apparent, but these are consistent with no-variation within $3\sigma$
error-bars, and hence, are less significant.} 55--160~s. The DM-range
sampled by these bursts (1.4--2.6~pc~cm$^{-3}$) corresponds to a maximum
change of 0.005~cm$^{-3}$ in the corresponding average electron density
($\langle n_e \rangle$) of the intervening medium across 250~pc. Such a
change in ISM is possible, albeit only at very large timescales (of the
order of years or decades). A larger change in electron density of a
correspondingly smaller portion of the medium could be a more likely reason
for the observed changes in DM. For example, if our sightline passes through
a medium slab of thickness d$_{\rm pc}$ parsecs entirely responsible for the
observed variation of 1.2~pc~cm$^{-3}$ in DM, then the required change in
electron density of this slab would be 1.2/d$_{\rm pc}$.
Given the variation timescale and pulsar's transverse velocity of nearly
210~km~s$^{-1}$, even a slab existing just next to the pulsar would imply
appropriate thermal electron density variations to occur over extremely
small spatial scales of just $\approx$10000~km (i.e., $\sim$10$^{-9}$~pc).
Moreover, the required change in electron density at such a
spatial scale would be of the order of $10^9$~cm$^{-3}$. Such changes, at
the required spatial scales, are unlikely to happen in the intervening
medium even very near the pulsar, but might be possible within the pulsar
magnetosphere. Possibility of a magnetospheric DM seems unlikely, although
minor contribution of the magnetosphere towards the observed RM has not been
excluded completely \citep{Noutsos09}.
Overall, the explanation for the observed short timescale change in DM would
require a radically new theoretical approach, and detections of more bursts
from the pulsar through extensive dedicated observations will help in obtaining
more clues for the same.
\section{Discussion and Conclusions}
Radio emission from Geminga pulsar has remained puzzling, with
comparable number of claims supporting the \emph{radio-quiet} and
\emph{radio-loud} nature of the pulsar. Claimed detections of pulsed
emission from the pulsar have been of low significance and the reported
flux density appears to be highly variable. Even if the claimed detections
are real, non-detections in several other sensitive searches suggest the
radio emission from the pulsar to be non-persistent and variable.
Although there has not been any other systematic search for transient
emission from the pulsar, extensive searches have been conducted for
periodic signal \citep[e.g.,][]{Ershov07}, and there has not been any
precedence of detecting strong pulsed emission from the pulsar. Even in
our 99 observing sessions, the
bursts have been detected only in 2 sessions. Hence, the emission of
such strong radio bursts from Geminga pulsar is a rare phenomenon.
\par
Energies of the bursts compare well with those of the giant-pulses
from the Crab pulsar at decametre wavelengths. The intrinsic widths
of the giant pulses are known to be extremely narrow (nanoseconds to
microseconds), compared to the moderately large widths (30--270~ms)
of the bursts. Given the low DMs, the observed widths are unlikely to
be associated with scattering in the intervening medium (in
the thin screen approximation). However, the large widths might have
contribution from the same mechanism that is responsible for the
observed short timescale variation in DM.
\par
\citet{HR93} have suggested that $\gamma$-ray emission from the
outer-magnetosphere could produce copious e$^{\pm}$~pairs in the
inner-magnetosphere to quench Geminga's radio emission. They also note that
for this quenching mechanism to work, a large inclination angle between the
magnetic and rotation axis (so that a significant fraction of $\gamma$-rays
pass through the inner-magnetosphere) is needed, and models of Geminga's
gamma-ray emission indeed suggest a significantly inclined dipole
\citep[e.g.,][estimate an inclination angle of $\sim 50$\mdeg, and a viewing
angle of $\sim 86$\mdeg]{ZC01}. On the basis of transient dips in Geminga's
soft X-ray profile, \citet{HW97} suggest that the supply of copious
e$^{\pm}$~pairs in the inner-magnetosphere may be variable, and hence,
occasional clearing away of the quenching plasma might imply Geminga to be
a {transient} radio emitter. If the claimed detections of radio emission
from Geminga pulsar have indeed benefitted due to occasional ceasing of the
quenching e$^{\pm}$~pairs, detection of strong radio bursts might imply a
complete cessation of the quenching plasma. Note that a small number of
e$^{\pm}$ pairs, e.g., just before cessation happens, might even help in
triggering the creation of plasma pairs in the polar gap required for
the radio emission process \citep{RS75}. Dedicated observations of the
pulsar simultaneously at radio frequencies and in X-rays can decisively
find out if the above quenching mechanism is responsible for Geminga's
radio quietness.
\par
The short timescale variation in DM, if real, can also explain
non-detection of periodic emission from Geminga pulsar in some of the
earlier sensitive radio searches. Periodicity search crucially depends
on the in-phase addition of pulsed signal. However, the variation in DM at
timescales shorter than the integration time would imply a corresponding
variation in arrival times of individual pulses. Hence, the short timescale
variation in DM would make the periodicity search inadequate, and a search
for a faint periodic signal using deep integration would be ineffective.
However, note that the above effect is prominent only at lower frequencies
($\lesssim 150$~MHz). Sensitive upper limits at higher radio frequencies
together with detection of radio bursts presented here, and possibly the
earlier claimed radio detections, indeed suggest Geminga pulsar to be
radio-loud only occasionally.
\par
Summarizing, we have presented discovery and detailed properties of
several fast radio transients, with some of them highly
linearly polarized, from the gamma-ray pulsar Geminga.
These bursts exhibit variation in their dispersion measures (DM) over
timescales as short as a minute, which
is hard to explain by corresponding change in the electron density of
the intervening medium. The short timescale variation in DM has remained
the biggest puzzle in our findings.
The energies of the bursts are comparable with those of giant pulses from
Crab pulsar at decametre wavelengths, and might actually suggest the bursts
to be radio giant pulses from Geminga. 
We have also discussed possibility of radio emission from Geminga due to
occasional cessation of the copious quenching plasma in the pulsar's
inner-magnetosphere \citep{HR93,HW97}. Detection of more bursts from
the pulsar through dedicated low frequency observations can provide
crucial clues to the underlying mechanism for occasional radio emission
from Geminga, and possibly also from other radio-quiet gamma-ray pulsars.
Discovery of these bursts have also demonstrated the potential of transient
searches at very low frequencies.
%
%
%
%
\section*{Acknowledgments}
I thank the anonymous referee for a thorough review, and useful comments and suggestions.
I am thankful to Aswathappa H. A. for help with the observations, and gratefully
acknowledge the support from the observatory staff. The Gauribidanur radio
telescope is jointly operated by the Raman Research Institute and the Indian
Institute of Astrophysics.
I am grateful to Roman Gorgutsa for providing archived solar monitoring data from
IZMIRAN's solar radio laboratory (\url{http://www.izmiran.ru/stp/lars/}), and
thankful to Indrajit Barve for help in obtaining these data. Use of
archived solar radio spectra provided by the Culgoora and Learmonth Solar Radio
Observatories (Western Australia) has also been made, and I gratefully acknowledge
Vasili Lobzin and Divya Oberoi for their help in obtaining these data. I am
grateful to Harsha Raichur, Nishant Singh and Divya Oberoi for useful discussions
and comments on the manuscript, and to Avinash A. Deshpande and Bhal Chandra Joshi
for useful suggestions at early stages of this work. Thanks are due to Arun K.
Naidu, Bhal Chandra Joshi, P. K. Manoharan and M. A. Krishnakumar for providing
the B0740$-$28 data acquired using the software backend of the Ooty radio telescope.
I am thankful to the members of the LAT-team for providing the up-to-date timing
models of several gamma-ray pulsars including Geminga.


\end{document}